# Sensory-driven microinterventions for improved health and wellbeing

Youssef Abdalla[1], Elia Gatti[2], Mine Orlu[1], and Marianna Obrist[2]

[1]UCL School of Pharmacy, Research Department of Pharmaceutics, University College London, London, United Kingdom

[2]Department of Computer Science, University College London, London, United Kingdom

*corresponding author email address: m.obrist@ucl.ac.uk

## Abstract

The five senses are gateways to our wellbeing and their decline is considered a significant public health challenge which is linked to multiple conditions that contribute significantly to morbidity and mortality. Modern technology, with its ubiquitous nature and fast data processing has the ability to leverage the power of the senses to transform our approach to day-to-day healthcare, with positive effects on our quality of life. Here, we introduce the idea of "Sensory-driven microinterventions" for preventive, personalised healthcare. Microinterventions are targeted, timely, minimally invasive strategies that seamlessly integrate into our daily life. This idea harnesses humans' sensory capabilities, leverages technological advances in sensory stimulation and real-time processing ability for 'sensing the senses'. The collection of sensory data from our continuous interaction with technology (e.g. tone of voice, gait movement, smart home behaviour) opens up a shift towards personalised technology-enabled, sensory-focused healthcare interventions, coupled with the potential of early detection and timely treatment of sensory deficits that can signal critical health insights, especially for neurodegenerative diseases such as Parkinson's disease.

## The senses, their functioning and relevance for health and wellbeing

In our daily lives, we continuously process sensory information. Our main senses - smell, sight, hearing, touch, and taste - are essential for daily functioning and health maintenance. They constantly gather cues and send them to the brain, where they are processed, enabling us to perceive and interact with our surroundings[1]. These senses do not operate in isolation, instead, they blend together to provide a richer representation of the world we live in[2]. As such, sensory experiences profoundly influence our behaviour, decisions, and emotions[3], and shape our memories. Imagine how your memories would be like, if you experience the world without seeing? Reduced or no vision and hearing impairments can significantly diminish everyday experiences, social interactions and overall quality of life[4]. Perhaps less apparent are the effects of dysfunctions on our sense of smell, taste, and touch. What would eating be like if you couldn't smell the flavours or savour the seasoning? Beyond impact on day-to-day interactions, disorders and the degradation of these senses also possess significant health and wellbeing challenges to those affected[5]. Interventions such as olfactory training have proven positive effects on olfactory functions[6], smell rehabilitation, and promise additional benefits to counteract the natural decline of the sense of smell in older age[7].

The early diagnosis of sensory decline can also support practitioners in the prompt identification of and intervention for several conditions that manifest as a decrease in sensory-perceptual abilities. For example, Transient Ischemic Attacks (TIA)[8] brain tumours[9], Multiple Sclerosis (MS)[10] and diabetes[11] can all manifest in their early stages through a vision deficit. Hearing, taste, and smell losses and disturbances can all be predictors of multiple neurological, psychiatric, and systemic diseases, such as dementia[12-14]. There is growing evidence of smell being an early biomarker for neurodegenerative disease development. The most common non-motor symptom in Parkinson's disease patients is smell (olfactory) impairment, occurring at least 4 years prior to motor symptom onset[17]. While we cannot avoid this disease developments, regular sensory assessments can enable early detection and diagnosis. Thus, offering the growing ageing population the opportunity for a longer independent living and timely treatments. This can start with self-care and an increased awareness about the senses in everyday life, such as illustrated in Figure 1.

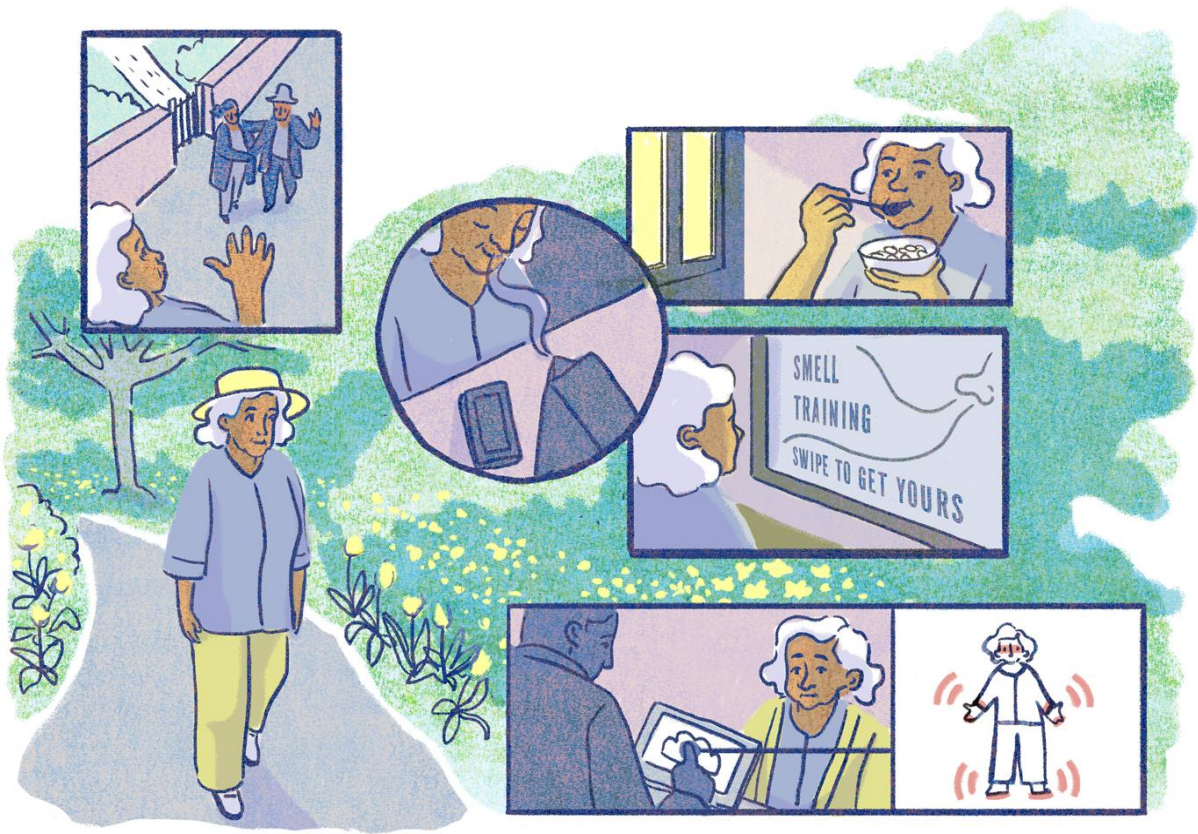

*Figure 1: Speculative scenario where regular smell training enabled an early identification of Parkinson's disease. Smell training is only an example of an early sensory stimulation approach that is relevant for the broader sensory-driven microintervention approach.* Illustration credit: Ana Marques.

*It is the year 2050. Emma, a retired artist, was diagnosed with Parkinson's disease ten years ago. While the disease is progressively affecting her movement, balance, memory and sensory perception, her quality of life isn't as bad as for her 76-year-old neighbour John, who relies on the social care system for many day-to-day tasks, from hygiene, cooking, to walking.*

*Emma's journey was different thanks to an innovative approach that had already been in use for years before her diagnosis. In fact, when Emma was 54 years old, she took part in a smell training study, where she received a digital scent-delivery device and App to keep track of her olfactory perception at home. While the initial trial ended after 6-months, she kept being more aware about her sense of smell and other sensory capabilities, mindfully eating and enjoying the flavours, as well as taking in the sounds, colours and aromas when walking through the park or along the beach. Five years later, when Emma saw an advertisement for the smell training system on TV, she bought a subscription and kept track of her perceptual capabilities, which was a good thing, as she noticed that over the span of two years here sense of smell got worse. Its normal to have a reduced smell as we age – she knew that – but there were some anomalies that felt wrong; why couldn't she enjoy her favour coffee anymore.*

*When she went to her GP, he looked at her healthcare records, which was with her permission also linked to her smell training data. The GP ran a smell test and ordered some blood tests, both confirming early signs for an increased risks for Parkinson's disease development. Emma was signed up for a new 'sensory-driven micro-intervention' approach, extending her existing smell training routine with new features, especially on movement, gait tracking, nutrition and*

*cognitive behaviours. The new system, allowed her to keep track of her sensory capabilities and behaviours automatically, embedded with tailored sensory training and whenever needed sensory substitution interventions to augment her day-to-day tasks. These targeted exercises have allowed her to maintain her independence, develop neural resilience and adaptability, transforming her experience of living with Parkinson's through a prolonged enhanced quality of life.*

In this future scenario the main character Emma was diagnosed with Parkinson's disease early on, enabled by regular self-monitoring of her olfactory functions, and the integration of her olfactory training data with her healthcare records. Despite growing scientific evidence on the importance of sensory data in relation to early diagnosis, detection and treatment of disease development, we still lack clear guidance on how to harness sensory-driven data for efficient health innovation. In this article, we offer a perspective on a future-looking preventive, personalised healthcare approach: ***sensory-driven microinterventions***. This approach is increasingly made feasible due to technological advancements in digital health innovations, including the emergence of sensory devices and interfaces that go beyond assessments of visual and hearing capabilities[15], as well as the undeniable potential of artificial intelligence (AI) and connected sensor and Internet of Things (IoT) systems. Those technologies offer real-time data collection and processing infrastructure, personalisation through machine learning algorithm, and a unique ecosystem for embedding sensory-driven microinterventions.

## The importance and effectiveness of microinterventions

Microinterventions refer to small, targeted actions or treatments of very short duration, often repeated over time, either with regular cadence or triggered by specific conditions in the patient, aimed at addressing specific health issues, improving patient outcomes, or enhancing the delivery of healthcare and mental health and wellbeing services[16,17]. Microinterventions vary in depth, duration, and timing of content delivery. In the form of few, short breathing exercises at strategic moments, they have been shown to help managing stress and increase mindfulness[18]. As video-guided gratitude and mindfulness tasks, microinterventions have been proven to increase body satisfaction[19]. As memory recall exercises executed before sleep, they have been shown to significantly decrease the occurrence of bad dreams in subjects[20]. People often unconsciously seek microinterventions in everyday situations, such as playing an uplifting song. It has been shown that music-based microinterventions can effectively reduce stress, and are associated with decreased physiological arousal, cortisol levels, and heart rate[21]. Microinterventions have shown considerable efficacy in psychology, particularly in enhancing mental health and fostering behavioural change[22].

Unlike broad, systemic interventions or major surgical procedures, microinterventions are often subtle, focused, and less invasive. They can play a significant role in preventative care, early treatment, and patient management, especially in chronic diseases or in improving patient experiences within healthcare systems[23]. Specific types of microinterventions, called 'Just-in-Time Adaptive Interventions' involve resources that are quick to consume and are designed to elicit an immediate positive response, aligned with moments of greatest risk[24]. Other types of microinterventions are less strict when it comes to the timing of the intervention delivery[19], but still maintain their easy-to-access and quick-to-consume characteristics. Despite this promising work, microinterventions remain under-researched in healthcare and wellbeing, and do not leverage the potential inherent in sensory interaction and perception.

## The potential of 'sensory-driven' microinterventions in digital health

Beyond traditional microintervention approaches, we reflect here on the opportunities sensory-driven microinterventions can offer for improved health and wellbeing outcomes. For example, sensory training and rehabilitation can substantially enhance the quality of life of patients suffering from a myriad of conditions[25]. Sensory-impaired populations are more prevalent than one might think. For example, 5% of people are thought to have anosmia, the inability to smell, with around 20% facing some form of smell disorder[26]. This number rises to 75% for people aged between 70–80 years[26,27]. With the increasingly ageing world population, the number of people affected by sensory deficits continues to increase. Taste, smell and tactile sensitivity[28], proprioception[29], vision[30] and hearing[31] all deteriorate with age. Most importantly, these sensory deficits contribute to morbidity, disability, and mortality[32]. Given their profound impact on health and wellbeing, the timely identification of any deterioration in sensory perception and intervening, where possible, to slow the decline can have a profound impact on people's lives.

In clinical practice, many treatments consist of delivering sensory stimulation to patients, divided into 3 main categories: sensory training, sensory substitution, and sensory integration therapy. In *sensory training*, sensory stimuli can be used to restore perceptual abilities that have been damaged by ageing or underlying diseases. For example, delivering prolonged mechanical noise in neuropathic patients has been shown to improve their ability to perceive tactile stimuli[25], in turn improving their motor skills[33] and posture[34]. Furthermore, sensory training has been exploited to improve the action-perception loop[35] by providing exercises focused on strengthening proprioception[36] visual processing[37], and balance[38]. Finally, it is important to highlight that, while the body of work on training chemical senses is somewhat less developed, smell and taste have also been trained with success in healthy adults, improving stimuli discrimination and recognition[7,39]. For example, smell training using odours can improve olfactory functioning[6] and digital solutions are developed to enable self-monitoring at home[7].

Next to sensory training, a single sense may not be as effective as providing complementary information through another sensory modality. Hence, the second treatment category refers to *sensory substitution* that acts by replacing or complementing information typically conveyed by one sensory modality to another, for example based on correspondences between vision through sound[40]. It is widely known that sensory substitution has been used in translating visual stimuli to acoustic and tactile ones, to aid visually impaired subjects in reading tasks[41]. Recently, sensory substitution has become more reliant on technology, with systems detecting and analysing stimuli from one sense and translating them into another[42]. For example, technology-powered sensory substitution systems can now use tactile stimuli to replicate pictures[43] or use spatialised sound to enhance 3D perception[43]. Sensory substitution has also been used to provide compelling musical experiences for individuals with hearing dysfunctions through exploring the use of colours and haptics to "translate" musical experiences[44] while other researchers use vibrations to "feel" musical performances[45].

Finally, sensory stimuli can be delivered as part of a treatment for many developmental, neurological, and psychological conditions, in what has been called sensory therapy or *sensory integration therapy* (SIT)[46]. SIT can take various forms, delivering different sensory stimuli across different sensory modalities, often to enhance attention[47], helping to manage emotional reactions[48] and improve social skills[49]. In fact, sensory therapy strategies fit the short delivery, easy-to-consume paradigm which is the trademark of microinterventions. Songs, which only last a few minutes, have been used in the past as guise of microinterventions to manage stress[21] and several studies showed how just a few minutes of tactile stimulation can deliver benefits in managing emotions to patients of different age groups, although the benefits may be reduced compared to prolonged stimulation[50]. Notably, almost all the microinterventions delivering sensory stimuli as therapy act towards the psychological wellbeing of the users. In other words, microinterventions that deliver sensory stimuli to users heavily focus on "sensory integration therapy", with no examples of using microinterventions for "sensory training" and "sensory substitution".

This article puts an emphasis on those often overlooked and neglected treatment approaches for preventive and personalised health care. Technological advancements increasingly allow us to reflect upon and speculate about the future opportunities of detecting patterns that stem from deviations in sensory and perceptual capabilities, as well as for delivering short, timely stimulation to tackle sensory deficits either through training or sensory substitution. This speculations are grounded in the growing ability of "sensing the senses" and "intervening for the senses" through technological advancements enabling digital microinterventions[51,52].

## Microinterventions for sensory training and substitution

The idea of sensory-driven microinterventions embedded in everyday life and to deliver timely and relevant sensory stimulation to benefit individuals' healthcare and wellbeing, as well as provide healthcare professionals with timely and up-to-date patient information. Notably, those microintervention can deliver personalised sensory stimulation for training and substitution accounting for personal preferences, variation in perceptual thresholds, and aligned with daily routines to generate bespoke sensory protocols for each individual and to maximise health and wellbeing outcomes. For example, hearing impairments could be picked up by virtual assistants analysing the voice of the users[53]. Smart home sensors and phone data could then be used to analyse the user's day and design a suitable sensory training regime to slow down hearing loss that fits the user's schedule (e.g. integrated with daily meditation sessions). The system could also alert users when a visit to a healthcare professional if necessary. Additionally, the assigned healthcare professional could be notified to review any anomalies in the data and determine whether an in-person visit or continued remote monitoring with further data collection would be more beneficial. This approach could help reduce the strain on healthcare systems, lower costs, and conserve increasingly stretched resources.

If we now think back to the future scenario of Emma (see Figure 1), who was diagnosed with Parkinson's disease early on, how can we imagine her living with Parkinson's considering the benefits of the sensory-driven microintervention treatment she was enrolled? Here is what we collaboratively imagined as a desirable future – a day in the life of Emma (see Figure 2).

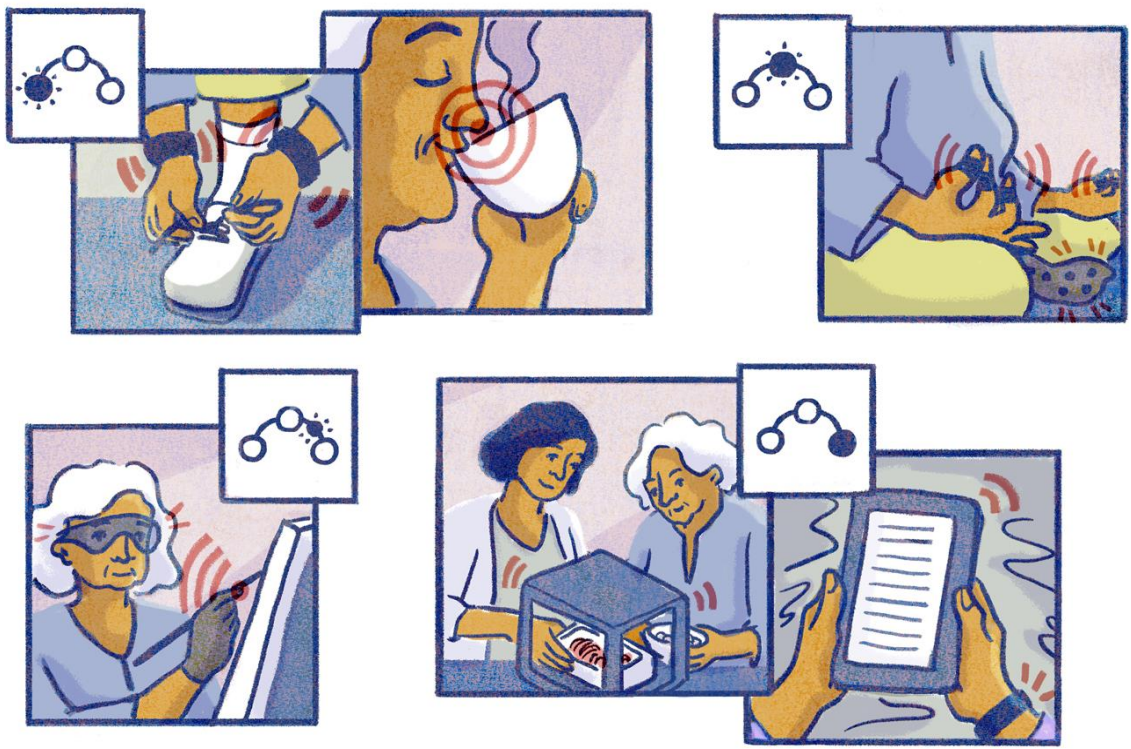

*Figure 1: A Day in the Life of Emma, who is living with Parkinson's and the envisaged benefits of early sensory-driven microintervention enabled through technological advancements from wearables, sensory stimulation devices/interfaces, food printing and smart sensing/IoT systems.* Illustration credit: Ana Marques.

*Emma wakes up to the soft chime of her bedside alarm. As she gets dressed, her adaptive smart wristband gently vibrates, helping to stabilize her tremors. These small yet significant adjustments make routine actions smoother, allowing her to start her day with confidence. Dressing has become easier over time. As she applies gentle pressure to the laces, the device converts it into soft auditory tones, guiding her hands with precision. The once-dreaded task of tying her shoelaces is now seamless, thanks to her smart adaptive sensory substitution*

*system that also records and monitors her tactile acuity and modifies the pressure profile as needed.*

*In the kitchen, Emma prepares her breakfast, savoring the decision of selecting her morning coffee. Her years of smell training, or as she refers to as 'nose gym', have refined her ability to differentiate subtle aromas, turning this daily choice into a memory-enhancing exercise. With seven distinct coffee brews in her kitchen, each linked to a specific day, she enjoys a cognitive boost simply by engaging her senses.*

*At midday, Emma takes a moment for mindfulness meditation, sitting comfortably in her favorite cushion. Gentle stimulations on her fingers and toes create subtle tactile sensations, enhancing her perception of touch and grounding her in the present. The rhythmic pulses guide her breathing, reinforcing a sense of calm and stability. As she focuses on each sensation, she becomes more aware of her body's connection to her surroundings, reducing anxiety and promoting relaxation.*

*Emma, remembering her time as a young artist, stands at her easel, picking up her paintbrush with renewed excitement. With her color-enhancing glasses compensating for her declining visual acuity, she experiences the vibrancy of each shade with clarity. Each vibration is also translated into a musical expression, enhancing not only her performance but also substituting some of the lost tactile sensations through sound. Art, once a struggle, has again become a source of joy and self-expression, even if different than her younger self.*

*For dinner, Emma is joined by her daughter Mira. While they catch up on gossips from last week, the food printing module started crafting a meal based on Emma's flavour profile, but having detected Mira's arrival as 'approved user' earlier, her flavor preferences are also considered in crafting the dishes. To enhance their dining experience, sonic seasoning subtly enriches Emma's perception of texture and taste, making every bite more satisfying.*

*As Emma winds down, she picks up her e-reader, which has adapted to her needs with adjustable font sizes and tactile feedback. Subtle vibrations signal when it's time to rest. Her wristband begins recording vital data while triggering a gentle scent release designed to promote deep, restful sleep. With these small but powerful interventions, Emma drifts off, comforted by the knowledge that she is supported in her journey—one sensory moment at a time*

This scenario depicts one day in the life of Emma. It is showing a collaborative speculation of a desirable future when living with Parkinson's disease, enabled through technological advancements and the integration of sensory-driven microintervention. The success of sensory-driven microintervention, as imagined in this future scenario (Figure 2) relies on technologies in two major ways: *First* for 'sensing the senses', the use embedded sensors and users' interaction and behavioural data to assess the users' sensory perceptual capabilities and, when needed, to design a pattern of interventions that integrates seamlessly with the users' day-to-day activities and known sensory profile (continuously updated through new data points). *Second* for 'intervening for the senses', to deliver the sensory-driven micro-intervention stimulation for timely, personalised sensory training or sensory substitution. This technology could be as widespread as audio devices like headphones or earbuds, vibration patterns in smart devices, but could also be considerably more complex when involving other senses such as smell and taste, where sensory devices and interfaces are only emerging[54]. Those technologies taken together form the foundation for a personalised, continuous monitoring network and ecosystem, updated and adapted to a users' sensory profile and training and substitution patterns and practices.

## Insights from 'sensing the senses'

Advances in technologies such as wearables, IoT, and AI offer a timely opportunity to create a holistic, sensory-driven approach to personalised and preventive healthcare. However, these technologies currently work in isolation, limiting their potential impact. To efficiently and meaningfully deliver short, targeted, and personalised sensory-driven microinterventions, we need to connect those multiple data sources to create a reliable profile of users' perceptual and sensory abilities and interpretation of the recorded users' behaviours and patterns over time.

For example, sensory changes, such as in a person's voice recorded through their interaction with a smart assistant, their phone, or any microphone-equipped device can provide information about their hearing abilities. Subsequently, recognition of the volume at which they speak could indicate a loss of hearing. Similarly, an increase in ambient lighting could be interpreted as a decrease in visual acuity; or a change in the gait recorded from a phone's accelerometer might suggest issues with a person's balance and repeatedly dropping objects might suggest sensorimotor issues of various natures, including tactile, proprioceptive, and visual deficits. In a future smart home scenario, recordings of its smoke alarm going off repeatedly might suggest the presence of anosmia (loss of the sense of smell), potentially signalling early neurodegenerative diseases development like Alzheimer's and Parkinson's, more prevalent in older age, but also an early warning sign for other associated health and mental health issues.

These examples are only a snapshot of the data generated today through our constant interaction with technology. It is important to remember that behavioural changes related to the senses and sensory perception can have many causes. One might speak louder because of excitement, rather than because of hearing loss. Brightness could be increased in a room because something got lost and a higher than usual visibility is needed. The collection of sensory-related data therefore cannot be considered in isolation nor for single instances. Such data will need to be integrated with longitudinal data encompassing the "sensorial history" of an individual, their known psychophysical condition, their transient context, and their preferences. Moreover, there has to be an option for the integration of users' subjective assessments and feedback, considering the highly subjective nature of sensory experiences and perceptual variability across and within people over time[55]. Such self-reporting features may need to be kept to a minimum and strategically placed throughout time to reduce the demand for the individual and yet offer both control and agency concerning their own health and wellbeing data.

## Ability to 'intervening for the senses'

Alongside the identification of sensory changes and monitoring those changes over time, sensory-driven microinterventions depend on future technological innovations in the domain of sensory devices and interfaces to capture the full spectrum of humans' sensory abilities. Devices for olfactory and gustatory stimulation are emerging[54] and have the ability to deliver sensory stimuli for all the main senses in order to train sensory capabilities and substitute for sensory deficits.

Given the transient and in-the-moment nature of these envisioned sensory-driven microinterventions, digital solutions designed for sensory stimulation must meet specific conditions: (i) they should be portable, enabling stimulation to be delivered whenever required, often on short notice; (ii) they should be easy to use, as individuals receiving these interventions may have perceptual impairments and are likely to belong to the ageing population; and (iii) they should be adaptable and personalisable, as sensory deficits caused by cognitive impairments which can make it harder for individuals to maintain their devices.

These devices should also be capable of detecting even minor deviations from a person's baseline to ensure effective and timely adjustments and actions by healthcare professionals.

Today, most sensory stimulation devices that can deliver substitutions satisfying all these conditions have been built to aid visually impaired people in navigating their environment. The white cane, for example, represents a simple, durable, and portable device designed to effectively translate environmental, typically visual, information to haptic stimulation[56]. However, we also see chemo-sensory interfaces such as a tongue display that satisfies the portability, ease-of-use, and resilience characteristics[57] and uses electrodermal patterns delivered on the tongue to aid users' navigation.

Traditionally, most devices substituting the sense of sight use vibrotactile feedback to communicate the presence and distance of obstacles during navigation[58]. Sensory substitution devices delivering information supplementing senses other than vision are rarer and often less portable. Virtual reality (VR) headsets can be used to supply additional visual information to support users that are hard of hearing[59] and ad-hoc vibrotactile devices have been created to deliver localised information that conveys auditory experiences[60], provide information about the environment to visually impaired users[61], and allow for a better signal discrimination in noisy environments for hard of hearing people[62].

On the other hand, sensory training has traditionally not relied on specific devices to deliver sensory stimulation. Instead, interaction paradigms leveraging general-purpose devices, such as computers or smartphones, have been developed over time. For instance, video game-based training has been proposed as a potential method to enhance visual sensitivity and attention in users[63]. Similarly, computer applications have been utilised to improve auditory perception, often focusing on enhancing phoneme perception and discrimination[64]. In contrast, tactile sensory training has typically avoided the use of specific devices, favouring physical props such as real objects or textured surfaces[65]. A notable exception to this trend is smell training, where the controlled delivery of odours required for odour discrimination and detection has been achieved using specialised devices, digitally controlled by an app[17].

## Opportunities and grand challenges

Sensory-driven microinterventions have the potential to transform healthcare practice and improve the wellbeing of thousands, alleviating pressure on healthcare systems while enhancing the quality of life for a significant portion of the population. This is particularly true as challenges related to their execution are increasingly addressed through advancements in wearable and sensor systems, cloud computing, and AI/machine learning algorithms. These technologies, when embedded in real-world environments, effectively navigate their inherent complexities and adapt to dynamic conditions of everyday life.

At the same time, this approach, which leverages the human senses, offers a unique opportunity to advance technology and science, as it is not without its challenges.

***Challenge 1 Sensory profiling:*** One challenge when deploying sensory-driven microinterventions is the identification of sensory deficits through pattern recognition and analysis of a user's sensory history – an intriguing problem for machine and deep learning practitioners. This issue aligns with existing research in healthcare technology, where anomaly detection in user behaviour has been a focus for years. Techniques such as long short-term memory neural networks (LSTMs) and autoencoders have been widely explored to identify irregular behaviours[66]. If these methods are adapted to sensory data – which presents challenges in acquisition, storage, and selection – they could play a critical role in ensuring the accurate and timely delivery of sensory microinterventions.

***Challenge 2 In-the-moment delivery:*** Alongside the identification of sensory impairments, a significant challenge lies in developing effective delivery devices for sensory stimulation during microinterventions. While current technologies, such as mobile phones, are promising mediums for delivering vibrotactile, acoustic, and visual stimuli, the creation of bespoke devices tailored to optimise specific sensory training or sensory substitution paradigms involving smell and taste remains in its infancy. The development of feasible, compelling, and portable solutions for these chemical senses is still evolving. Notable progress has been made, particularly within the Human-Computer Interaction (HCI) research community, which has shifted from lab-based and clinical equipment toward more portable and adaptive systems, such as those designed for smell[15] and even taste[67].

**Challenge 3 Intervention design:** A challenge that is less technology-focused but more healthcare-oriented lies in the development of the microinterventions themselves. Researchers will need to design effective microinterventions that maximise the benefits and durability of sensory training and sensory substitution. This task merges technical challenges with the real-world need for meaningful and impactful interventions, requiring a fundamentally user-centric approach to address individual circumstances, needs, and preferences. As such, the creation of microinterventions will represent a cross-disciplinary challenge, demanding collaboration across fields to ensure their success.

**Challenge 4 From innovation to adoption:** Beyond technological and scientific challenges, cultural and logistical barriers must also be addressed. The acceptability of microinterventions both by the wider public and healthcare professionals will likely depend on the perceived security of the monitored data used to identify sensory decline. Adherence to sensory training programs will hinge on the usability of the devices and the thoughtful design of training protocols. Critical questions about data integrity, control over data flow, and access must also be considered when designing the ecosystem for sensory-driven microinterventions. Beyond simple biomedical models, a broader lifecourse-informed model to healthcare should be considered, fostering the 'preventive' approach as we also tried to illustrate in our speculative future scenario (Figure 1 and 2).

## Concluding remarks

In the current era, the widespread availability of digital technologies has significantly simplified the application of digital microinterventions. For instance, smartphone applications enable the timely delivery of interventions, addressing critical needs at pivotal moments[51], while web-based platforms have increased treatment accessibility at a lower cost[52]. Furthermore, with the continued advancement of AI, IoT, wearable sensor and sensory systems, and their increasing integration into daily life, sensory-driven microinterventions can seamlessly become part of everyday routines. This empowers individuals to fully harness their senses to achieve specific health and wellbeing goals and healthcare professionals to keep track of any anomalies early on. We envision a future where our senses are sharpened, trained, and unlocked to transform health strategies, enhance wellbeing, and improve quality of life outcomes as we age.

# References

1. Shams, L. & Beierholm, U. Humans' multisensory perception, from integration to segregation, follows Bayesian inference. *Sensory cue integration* **251** (2011).
2. Spence, C. Scent and the Cinema. *i-Perception* **11**, 2041669520969710 (2020). https://doi.org:10.1177/2041669520969710
3. Beard, J. R. *et al.* The World report on ageing and health: a policy framework for healthy ageing. *Lancet* **387**, 2145-2154 (2016). https://doi.org:10.1016/s0140-6736(15)00516-4
4. Kwon, H. J., Kim, J. S., Kim, Y. J., Kwon, S. J. & Yu, J. N. Sensory Impairment and Health-Related Quality of Life. *Iran J Public Health* **44**, 772-782 (2015).
5. Erskine, S. E. & Philpott, C. M. An unmet need: Patients with smell and taste disorders. *Clin Otolaryngol* **45**, 197-203 (2020). https://doi.org:10.1111/coa.13484
6. Hummel, T. *et al.* Effects of olfactory training in patients with olfactory loss. *The Laryngoscope* **119**, 496-499 (2009). https://doi.org:https://doi.org/10.1002/lary.20101
7. Beşevli, C. *et al.* Nose Gym: An Interactive Smell Training Solution. *Extended Abstracts of the 2023 CHI Conference on Human Factors in Computing Systems*, Article 464 (2023). https://doi.org:10.1145/3544549.3583906
8. Kim, J. S. Symptoms of transient ischemic attack. *Front Neurol Neurosci* **33**, 82-102 (2014). https://doi.org:10.1159/000351905
9. Jariyakosol, S. & Peragallo, J. H. The Effects of Primary Brain Tumors on Vision and Quality of Life in Pediatric Patients. *Semin Neurol* **35**, 587-598 (2015). https://doi.org:10.1055/s-0035-1563571
10. Sakai, R. E., Feller, D. J., Galetta, K. M., Galetta, S. L. & Balcer, L. J. Vision in multiple sclerosis: the story, structure-function correlations, and models for neuroprotection. *J Neuroophthalmol* **31**, 362-373 (2011). https://doi.org:10.1097/WNO.0b013e318238937f
11. Kovacova, A. & Shotliff, K. Eye problems in people with diabetes: more than just diabetic retinopathy. *Practical Diabetes* **39**, 34-39a (2022). https://doi.org:https://doi.org/10.1002/pdi.2378
12. Desai, N., Maggioni, E., Obrist, M. & Orlu, M. Scent-delivery devices as a digital healthcare tool for olfactory training: A pilot focus group study in Parkinson's disease patients. *Digit Health* **8**, 20552076221129061 (2022). https://doi.org:10.1177/20552076221129061
13. Li, S. *et al.* Hearing Loss in Neurological Disorders. *Front Cell Dev Biol* **9**, 716300 (2021). https://doi.org:10.3389/fcell.2021.716300
14. Langdon, R., McGuire, J., Stevenson, R. & Catts, S. V. Clinical correlates of olfactory hallucinations in schizophrenia. *British Journal of Clinical Psychology* **50**, 145-163 (2011). https://doi.org:10.1348/014466510X500837
15. Hopper, R. *et al.* Multi-channel portable odor delivery device for self-administered and rapid smell testing. *Communications Engineering* **3**, 141 (2024). https://doi.org:10.1038/s44172-024-00286-1
16. Paredes, P. *et al.* in *Proceedings of the 8th international conference on pervasive computing technologies for healthcare.*  109-117.
17. Meinlschmidt, G. *et al.* Smartphone-based psychotherapeutic micro-interventions to improve mood in a real-world setting. *Frontiers in psychology* **7**, 1112 (2016).

18   Elefant, A. B., Contreras, O., Muñoz, R. F., Bunge, E. L. & Leykin, Y. Microinterventions produce immediate but not lasting benefits in mood and distress. *Internet Interventions* **10**, 17-22 (2017). https://doi.org:https://doi.org/10.1016/j.invent.2017.08.004
19   Fuller-Tyszkiewicz, M. *et al.* A randomized trial exploring mindfulness and gratitude exercises as eHealth-based micro-interventions for improving body satisfaction. *Computers in Human Behavior* **95**, 58-65 (2019). https://doi.org:10.1016/j.chb.2019.01.028
20   Malouff, J. & Johnson, C. Effects of a micro-intervention aimed at reducing the level of unpleasant dreams. *International Journal of Dream Research*, 115-118 (2020).
21   de Witte, M., Knapen, A., Stams, G.-J., Moonen, X. & Hooren, S. v. Development of a music therapy micro-intervention for stress reduction. *The Arts in Psychotherapy* **77**, 101872 (2022). https://doi.org:https://doi.org/10.1016/j.aip.2021.101872
22   Kattenberg, K. *Supporting nurses' daily self-regulated learning behaviour via an online micro-intervention*, University of Twente, (2021).
23   Entenberg, G. A. *et al.* AI-based chatbot micro-intervention for parents: Meaningful engagement, learning, and efficacy. *Frontiers in Psychiatry* **14** (2023). https://doi.org:10.3389/fpsyt.2023.1080770
24   Nahum-Shani, I. *et al.* Just-in-Time Adaptive Interventions (JITAIs) in Mobile Health: Key Components and Design Principles for Ongoing Health Behavior Support. *Annals of Behavioral Medicine* **52**, 446-462 (2017). https://doi.org:10.1007/s12160-016-9830-8
25   Cloutier, R. *et al.* Prolonged mechanical noise restores tactile sense in diabetic neuropathic patients. *Int J Low Extrem Wounds* **8**, 6-10 (2009). https://doi.org:10.1177/1534734608330522
26   Desiato, V. M. *et al.* The Prevalence of Olfactory Dysfunction in the General Population: A Systematic Review and Meta-analysis. *Am J Rhinol Allergy* **35**, 195-205 (2021). https://doi.org:10.1177/1945892420946254
27   Schlosser, R. J. *et al.* A Community-Based Study on the Prevalence of Olfactory Dysfunction. *Am J Rhinol Allergy* **34**, 661-670 (2020). https://doi.org:10.1177/1945892420922771
28   Thornbury, J. M. & Mistretta, C. M. Tactile sensitivity as a function of age. *J Gerontol* **36**, 34-39 (1981). https://doi.org:10.1093/geronj/36.1.34
29   Shaffer, S. W. & Harrison, A. L. Aging of the somatosensory system: a translational perspective. *Phys Ther* **87**, 193-207 (2007). https://doi.org:10.2522/ptj.20060083
30   Wormald, R. P., Wright, L. A., Courtney, P., Beaumont, B. & Haines, A. P. Visual problems in the elderly population and implications for services. *Bmj* **304**, 1226-1229 (1992). https://doi.org:10.1136/bmj.304.6836.1226
31   Liu, X. Z. & Yan, D. Ageing and hearing loss. *J Pathol* **211**, 188-197 (2007). https://doi.org:10.1002/path.2102
32   Cross, H., Armitage, C. J., Dawes, P., Leroi, I. & Millman, R. E. "We're just winging it". Identifying targets for intervention to improve the provision of hearing support for residents living with dementia in long-term care: an interview study with care staff. *Disability and Rehabilitation*, 1-11  https://doi.org:10.1080/09638288.2023.2245746
33   Ochoa, N., Gogola, G. R. & Gorniak, S. L. Contribution of tactile dysfunction to manual motor dysfunction in type II diabetes. *Muscle & nerve* **54**, 895-902 (2016).

34    Simoneau, G. G., Ulbrecht, J. S., Derr, J. A. & Cavanagh, P. R. Role of somatosensory input in the control of human posture. *Gait & posture* **3**, 115-122 (1995).
35    Hurley, S. Perception and action: Alternative views. *Synthese* **129**, 3-40 (2001).
36    Aman, J. E., Elangovan, N., Yeh, I.-L. & Konczak, J. The effectiveness of proprioceptive training for improving motor function: a systematic review. *Frontiers in human neuroscience* **8**, 1075 (2015).
37    Appelbaum, L. G., Lu, Y., Khanna, R. & Detwiler, K. R. The effects of sports vision training on sensorimotor abilities in collegiate softball athletes. *Athletic Training & Sports Health Care* **8**, 154-163 (2016).
38    Runge, M., Rehfeld, G. & Resnicek, E. Balance training and exercise in geriatric patients. *J Musculoskelet Neuronal Interact* **1**, 61-65 (2000).
39    Hamilton-Fletcher, G., Alvarez, J., Obrist, M. & Ward, J. SoundSight: a mobile sensory substitution device that sonifies colour, distance, and temperature. *Journal on Multimodal User Interfaces* **16**, 107-123 (2022). https://doi.org:10.1007/s12193-021-00376-w
40    Macpherson, F. Sensory substitution and augmentation: An introduction.  (2018).
41    Deroy, O. & Auvray, M. Reading the world through the skin and ears: A new perspective on sensory substitution. *Frontiers in psychology* **3**, 457 (2012).
42    Lenay, C., Canu, S. & Villon, P. in *Proceedings Second International Conference on Cognitive Technology Humanizing the Information Age.*  44-53.
43    Bach-y-Rita, P. Tactile sensory substitution studies. *Ann N Y Acad Sci* **1013**, 83-91 (2004). https://doi.org:10.1196/annals.1305.006
44    Nanayakkara, S., Taylor, E., Wyse, L. & Ong, S. H. An enhanced musical experience for the deaf: design and evaluation of a music display and a haptic chair. *Proceedings of the SIGCHI Conference on Human Factors in Computing Systems*, 337–346 (2009). https://doi.org:10.1145/1518701.1518756
45    Frid, E. & Lindetorp, H. in *Sound and Music Computing Conferencem Torino, June 24th-26th 2020.*  68-75.
46    Smith, T., Mruzek, D. W. & Mozingo, D. in *Controversial therapies for autism and intellectual disabilities*    247-269 (Routledge, 2015).
47    Oh, H. & Kim, K. Effect of a multi-sensory play therapy program on the attention and learning of children with ADHD. *Journal of The Korean Society of Integrative Medicine* **7**, 23-32 (2019).
48    Rodriguez, M. & Kross, E. Sensory emotion regulation. *Trends in Cognitive Sciences* **27**, 379-390 (2023). https://doi.org:https://doi.org/10.1016/j.tics.2023.01.008
49    Khodabakhshi, M. K., Malekpour, M. & Abedi, A. The effect of sensory integration therapy on social interactions and sensory and motor performance in children with autism. *Iranian Journal of Cognition and Education* **1**, 39-53 (2014).
50    Packheiser, J. *et al.* A systematic review and multivariate meta-analysis of the physical and mental health benefits of touch interventions. *Nature Human Behaviour*, 1-20 (2024).
51    Baumel, A., Fleming, T. & Schueller, S. M. Digital Micro Interventions for Behavioral and Mental Health Gains: Core Components and Conceptualization of Digital Micro Intervention Care. *J Med Internet Res* **22**, e20631 (2020). https://doi.org:10.2196/20631

52　Muñoz, R. F. *et al.* Massive Open Online Interventions: A Novel Model for Delivering Behavioral-Health Services Worldwide. *Clinical Psychological Science* **4**, 194-205 (2015). https://doi.org:10.1177/2167702615583840
53　Glasser, A. T., Kushalnagar, K. R. & Kushalnagar, R. S. in *Proceedings of the 19th International ACM SIGACCESS Conference on Computers and Accessibility.* 373-374.
54　Cornelio, P., Vi, C. T., Brianza, G., Maggioni, E. & Obrist, M. in *Handbook of Human Computer Interaction* 1-31 (Springer, 2023).
55　Velasco, C. & Obrist, M. *Multisensory Experiences: Where the Senses Meet Technology*. (Oxford University Press, 2020).
56　Kristjánsson, Á. *et al.* Designing sensory-substitution devices: Principles, pitfalls and potential 1. *Restorative neurology and neuroscience* **34**, 769-787 (2016).
57　Kaczmarek, K. A. The tongue display unit (TDU) for electrotactile spatiotemporal pattern presentation. *Scientia Iranica* **18**, 1476-1485 (2011).
58　Cancar, L., Díaz, A., Barrientos, A., Travieso, D. & Jacobs, D. M. Tactile-sight: A sensory substitution device based on distance-related vibrotactile flow. *International Journal of Advanced Robotic Systems* **10**, 272 (2013).
59　Mirzaei, M., Kán, P. & Kaufmann, H. in *2021 IEEE Virtual Reality and 3D User Interfaces (VR).* 582-587 (IEEE).
60　Fletcher, M. D. Can haptic stimulation enhance music perception in hearing-impaired listeners? *Frontiers in Neuroscience* **15**, 723877 (2021).
61　Bach-y-Rita, P. Tactile sensory substitution studies. *Annals of the New York Academy of Sciences* **1013**, 83-91 (2004).
62　Cieśla, K. *et al.* Effects of training and using an audio-tactile sensory substitution device on speech-in-noise understanding. *Scientific Reports* **12**, 3206 (2022).
63　Green, C. S. & Bavelier, D. Effect of action video games on the spatial distribution of visuospatial attention. *Journal of experimental psychology: Human perception and performance* **32**, 1465 (2006).
64　Merzenich, M. M. *et al.* Temporal processing deficits of language-learning impaired children ameliorated by training. *Science* **271**, 77-81 (1996).
65　Carey, L. M., Matyas, T. A. & Oke, L. E. Sensory loss in stroke patients: effective training of tactile and proprioceptive discrimination. *Archives of physical medicine and rehabilitation* **74**, 602-611 (1993).
66　Miotto, R., Wang, F., Wang, S., Jiang, X. & Dudley, J. T. Deep learning for healthcare: review, opportunities and challenges. *Briefings in bioinformatics* **19**, 1236-1246 (2018).
67　Marzo, A. *et al.* Holographic acoustic elements for manipulation of levitated objects. *Nature Communications* **6**, 8661 (2015). https://doi.org:10.1038/ncomms9661